\documentclass[preprint,journal]{vgtc}       

\ifpdf
  \pdfoutput=1\relax                   
  \usepackage{graphicx}                
  \DeclareGraphicsExtensions{.pdf,.png,.jpg,.jpeg} 
\else
  \usepackage{graphicx}                
  \DeclareGraphicsExtensions{.eps}     
\fi%

\graphicspath{{figures/}{pictures/}{images/}{./}} 

\usepackage{microtype}                 
\PassOptionsToPackage{warn}{textcomp}  
\usepackage{textcomp}                  
\usepackage{mathptmx}                  
\usepackage{times}                     
\usepackage{cite}                      
\usepackage{tabu}                      
\usepackage{booktabs}                  

\usepackage{enumitem}
\setlistdepth{9}

\newlist{myEnumerate}{enumerate}{9}
\setlist[myEnumerate,1]{label=(\arabic*)}
\setlist[myEnumerate,2]{label=(\Roman*)}
\setlist[myEnumerate,3]{label=(\Alph*)}
\setlist[myEnumerate,4]{label=(\roman*)}
\setlist[myEnumerate,5]{label=(\alph*)}
\setlist[myEnumerate,6]{label=(\arabic*)}
\setlist[myEnumerate,7]{label=(\Roman*)}
\setlist[myEnumerate,8]{label=(\Alph*)}
\setlist[myEnumerate,9]{label=(\roman*)}

\usepackage{xspace}
\newcommand{\projectname}{\textit{CcNav}\xspace}

\ieeedoi{10.1109/TVCG.2020.3030357}

\onlineid{1203}

\vgtccategory{Research}
\vgtcpapertype{application/design study}

\title{CcNav: Understanding Compiler Optimizations in Binary Code}

\author{Sabin Devkota, Pascal Aschwanden, Adam Kunen, Matthew Legendre, and Katherine E. Isaacs}

\authorfooter{
\item
 Sabin Devkota and Katherine E. Isaacs are with University of Arizona. E-mail: \{devkotasabin@email.arizona.edu\,$|$\,kisaacs@cs.arizona.edu\}\,.
\item
 Pascal Aschwanden, Adam Kunen, and Matthew Legendre are with LLNL. E-mail: \{aschwanden1\,$|$\,kunen1\,$|$\,legendre1\}@llnl.gov\,.

}

\shortauthortitle{Devkota \MakeLowercase{\textit{et al.}}: CcNav: Understanding Compiler Optimizations in Binary Code}

\abstract{%

Program developers spend significant time on optimizing and tuning programs. During this iterative process, they apply optimizations, analyze the resulting code, and modify the compilation until they are satisfied. Understanding what the compiler did with the code is crucial to this process but is very time-consuming and labor-intensive. Users need to navigate through thousands of lines of binary code and correlate it to source code concepts to understand the results of the compilation and to identify optimizations. We present a design study in collaboration with program developers and performance analysts. Our collaborators work with various artifacts related to the program such as binary code, source code, control flow graphs, and call graphs. Through interviews, feedback, and pair-analytics sessions, we analyzed their tasks and workflow. Based on this task analysis and through a human-centric design process, we designed a visual analytics system Compilation Navigator (\projectname) to aid exploration of the effects of compiler optimizations on the program. \projectname provides a streamlined workflow and a unified context that integrates disparate artifacts. \projectname supports consistent interactions across all the artifacts making it easy to correlate binary code with source code concepts. \projectname enables users to navigate and filter large binary code to identify and summarize optimizations such as inlining, vectorization, loop unrolling, and code hoisting. We evaluate \projectname through guided sessions and semi-structured interviews. We reflect on our design process, particularly the immersive elements, and on the transferability of design studies through our experience with a previous design study on program analysis.

} 


\CCScatlist{ 
 \CCScat{K.6.1}{Management of Computing and Information Systems}%
{Project and People Management}{Life Cycle};
 \CCScat{K.7.m}{The Computing Profession}{Miscellaneous}{Ethics}
}

\teaser{
  \centering
  \includegraphics[width=\linewidth]{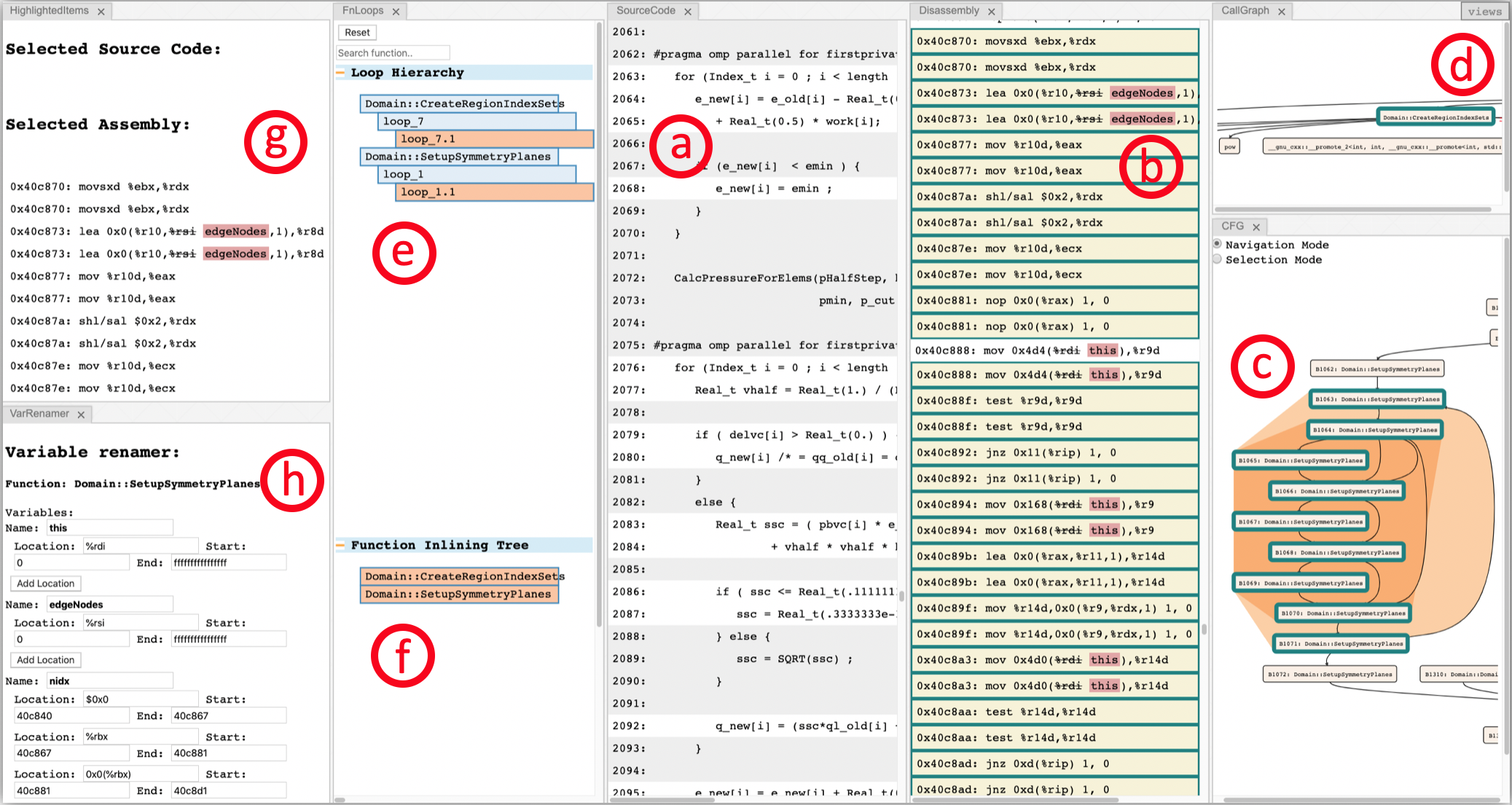}
  \caption{\projectname uses multiple coordinated views to enable correlation between source code (a) and disassembled binary code (b). A loop is selected in the loop hierarchy view (e). Matching disassembly is highlighted (b) with source variables annotated either with automated analysis or manual entry (h). The control flow graph (c), call graph (d), and function inlining (f) views provide extra context to the selection and alternative modes of navigation. A separate panel (g) collects all highlighted items for detailed examination.}
	\label{fig:teaser}
}

\vgtcinsertpkg

\begin{document}


\firstsection{Introduction}

\maketitle


\label{intro}

Demand for high performance computing (HPC) resources as well as scalability limitations of HPC applications drive the need for optimization. Even small percentage increases in efficiency can mean more science computed, either through more programs running or higher fidelity results than previously computationally possible. Thus, application developers and performance analysts spend significant time optimizing and tuning these programs.

One vector for optimization is at the compilation stage. When building the application, there are many choices in terms of which compiler to use and with what options. Furthermore, small non-algorithmic changes in the source code can lead the compiler to make different choices in how it transforms source code into machine-interpretable instructions. Running experiments can show which compiler version with which options performs better for a specific machine. However, for some applications performance is so critical that significant time and labor is devoted to trying to determine what optimizations were made by the compiler, whether they were effective, and what can be done to encourage it to optimize further. Understanding the optimizations may not only increase application efficiency on the target system, but lead to {\em portable} improvements where findings can be applied when compiling on another system.

Analyzing compiler optimizations is an iterative, experimental, and time-consuming task. Typically an analyst will disassemble a compiled binary into human-readable instructions and inspect in a text editor. They may also view the source code, make annotations, draw figures, and run ad hoc analyses. Even with debugging tools that show both source and disassembled code, analysts struggle to orient themselves in even moderately-sized programs of a few hundred lines of code.

This project is a collaboration between visualization and program analysis experts resulting in \projectname, a visual analytics tool to aid identification of compiler optimizations, their underlying causes, and their effect on performance. \projectname combines automated static analysis of compiled binaries with visual interfaces to support fine-grained analysis of compilation results. We conduct an ongoing design study~\cite{sedlmair-designstudies-2012} over 18 months with regular pair analytics~\cite{hernandez-2011-pair,elmqvist2012-pair} sessions and a three month immersive study. Through these activities we develop a data and task abstraction driving the design of our integrated system. We evaluate the system through pair analytics sessions with domain experts.

We find that experts in this style of program analysis employ a wide range of strategies, often jumping between whatever different abstractions and organizations of the data they had available to them and deriving or annotating new data. We therefore design \projectname to automatically derive views where possible, support linked navigation consistently through all views, and assist the most used forms of annotations. We also find that the collaborative and immersive nature of our meetings were fundamental in understanding these workflows.

We also reflect on the transferability of design studies based on our experience with a previous design study on program analysis~\cite{cfgexplorer-2018} that led the domain experts on our team to seek out the visualization experts. We describe the limitations of transferability, despite remaining in the same domain, and how our process either supported or dissuaded the transference of design.

In summary, our contributions are:

\vspace{-0.75ex}

\begin{itemize}
    \itemsep=0.25ex
    \item a data and task analysis for fine-grained analysis of compilation output (\autoref{sec:data}, \autoref{sec:hier-task-anls})
    \item the design and evaluation of a visual analytics system for analyzing the results of compilation (\autoref{sec:ccnav}, \autoref{sec:evaluation}), and
    \item a reflection on both transferability of visual solutions and immersive design techniques with implications for future visualization studies (\autoref{sec:reflection}).
\end{itemize}

\vspace{-0.75ex}

Before discussing these contributions, we provide a brief
overview of the domain and related work (\autoref{sec:background}). We then discuss our methodology in further detail (\autoref{sec:task_anls}). We conclude in \autoref{sec:conclusion}.



\section{Background and Related Work}
\label{sec:background}

Scientific simulation is used in diverse fields such as climate science, medicine, energy, and physics to study phenomena where it may be infeasible to do so otherwise. These simulations are frequently computationally-intensive and run on large, shared resources such as supercomputers and clusters. Thus, optimizing these programs to run faster can free resources for further scientific questions to be answered.

One avenue for optimizing these applications is to increase the efficiency of an algorithm through its translation for execution on a machine. \projectname aims to help developers and program analysts in understanding this process and ultimately using that understanding to generate more highly optimized software. We discuss the necessary background in compilation, optimization, and program analysis workflow, followed by a review of related literature.

\begin{figure}[tb]
 \centering 
 \includegraphics[width=\columnwidth]{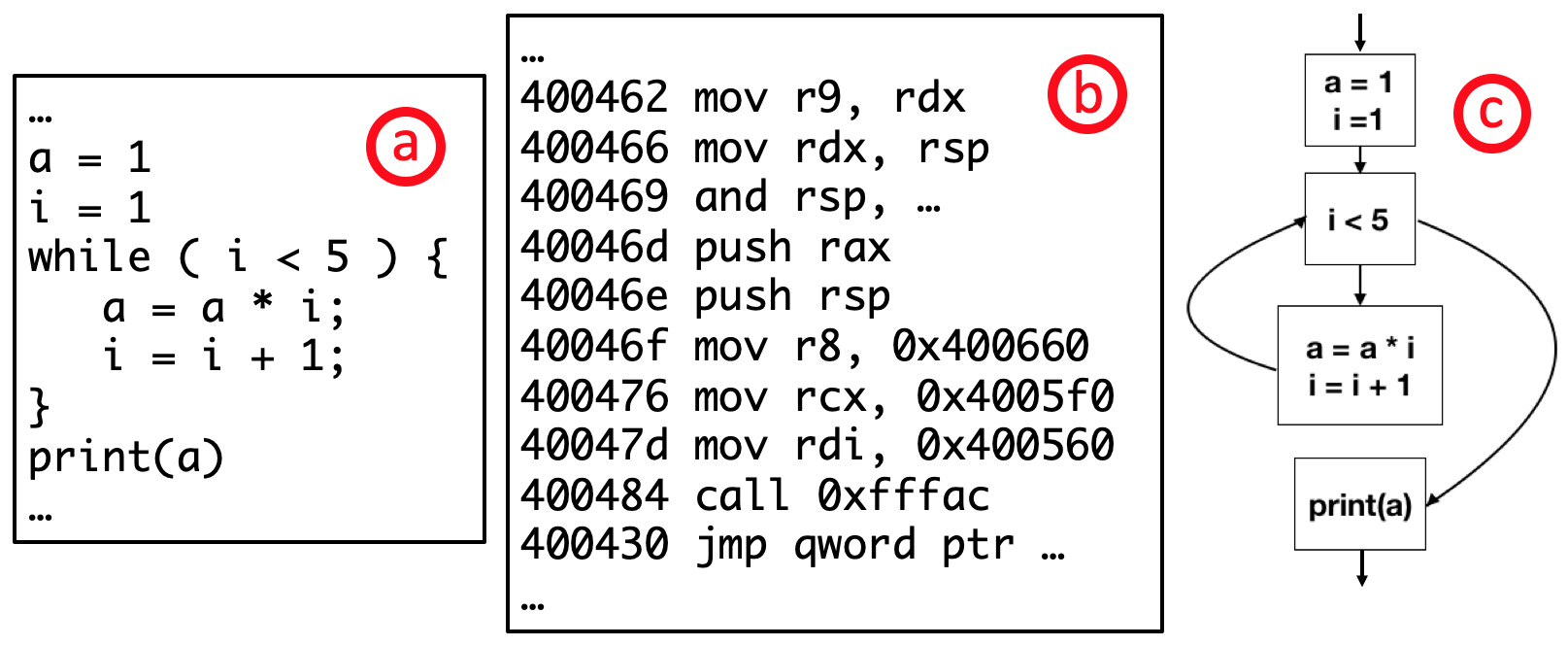}
 \caption{Source code (a), disassembly (b), and control flow graph (c).}
 \label{fig:bg_src_bin_cfg}
\end{figure}

\subsection{Compilation and Optimization}
\label{sec:compilation}

{\em Compilation} is the transformation of source code into a machine-interpretable format. The program that performs this transformation is known as the {\em compiler} and the resulting machine-executable software is known as the {\em executable} or {\em binary} file. As the machine-interpretable format is different between machines, this process must be done for each machine-architecture targeted. Thus, compilation is also an important element of code {\em portability} across machines.

People typically write source code in a higher-level language than that understood by the machine. As operations are available in the high level language that are not available to the machine, compilation is not a direct translation. The low level language of the machine is called the {\em instruction set architecture (ISA)} or {\em assembly language}. Assembly instructions typically have a format of an operation (e.g., \texttt{add}, \texttt{mov}) followed by parameters such as values or locations of values. These locations can be in memory or in temporary storage on the computation unit. These temporary locations are known as {\em registers}.

There are a multitude of valid transformations from source code to machine code. While the compiler must always generate correct code, it may also attempt to fulfill goals such as making the executable more efficient or producing a small binary. While in many contexts, developers are satisfied with the choices made by the compiler under default options, our collaborators are particularly concerned with generating more optimized code. Common optimizations seen in scientific code include function inlining, loop unrolling, and vectorization. Additionally, the compiler may create several {\em variants}---blocks of instructions that correspond to the same code but are optimized for different situations.

{\em Function inlining} removes the instructions (and therefore the overhead) of a function call by moving the body of the function within its calling function. This process sometimes requires duplicating instructions when a function is called from multiple places. {\em Loop unrolling} similarly removes overhead associated with checking loop bounds and jumping by placing several iterations of a loop body sequentially before performing the jump.

{\em Vectorization} translates repeated operations that might naively be performed in sequence to take advantage of parallel features of the computation unit. For example, a loop that multiplies every value in an array by a constant can be transformed to perform the operation in parallel across chunks of that array. The ISA typically has separate instructions and registers for vectorized operations.

To generate a more performant executable through compilation, developers can change the compiler (e.g., \texttt{gcc}, \texttt{clang}, or \texttt{llvm}), the compiler options (e.g., \texttt{-O3} for optimization-level-3 or \texttt{-funroll} to encourage loop unrolling), or even make small changes to the source code without changing the algorithm. However, even with these features, it can be difficult to predict what the compiler will do. 

Since performance is at a premium to our collaborators, they want to know whether the optimizations they expected were made and if not, what they can do to further encourage them. We call the collective strategies by which they answer these questions {\em program analysis}.

\subsection{Program Analysis}
\label{sec:programanalysis}

Developers can examine the results of the compilation by viewing the generated instructions, possibly with the help of automated tools. A compiled binary can be translated into human-readable machine instructions through a process known as {\em disassembly}. Typically the resulting text file is often referred to as {\em disassembly code} (or just `disassembly') and includes the {\em address} (memory location) associated with each instruction. These addresses are used to jump non-sequentially, e.g., in a loop or function call.

If the binary was compiled with an option to retain debug information, more information can be retrieved, such as mappings between source code and disassembly or whether a function was inlined. The quality of debug data is dependent on the features of the compiler. It is often incomplete or incorrect, especially in the presence of heavy optimization~\cite{YuanboLi2020PLDI}, so manual inspection is required.

To provide a structural interpretation of the disassembly, other structures, such as the control flow graph and call graph, may be derived from it. The {\em control flow graph (CFG)} divides the disassembly into {\em basic blocks}: contiguous address ranges that must be executed sequentially. Basic blocks are the nodes in the graph. Edges represent valid paths between basic blocks due to jumps, branches, and function calls. \autoref{fig:bg_src_bin_cfg} shows a small example. In the {\em call graph}, the functions are the nodes and the edges are valid calls between them. 

Dwarf~\cite{DWARF} is a popular format to support source level debugging. Objdump~\cite{objdump} and dwarfdump~\cite{dwarfdump} are popular tools for retrieving disassembly with debug information. Both produce text files. Dyninst~\cite{dyninst-2000} is a library for more advanced analysis. We use Dyninst as a basis for the automated analysis components of \projectname.


\vspace{1ex}

\noindent\textbf{Typical workflows.} Analysts typically use the above tools to get the disassembly and view both it and source code with a text editor, switching between views to orient themselves. They may also generate a CFG, sometimes filtered locally to the portion of the disassembly of interest. This is sometimes done with pen and paper or with tools like LLVM~\cite{LLVM} that generate a \texttt{DOT} file for rendering with the GraphViz \texttt{dot} algorithm~\cite{graphviz}. Our domain experts' interest in the CFGExplorer~\cite{cfgexplorer-2018} visualization over \texttt{dot} was an impetus for our collaboration.

\begin{figure}[tb]
 \centering 
 \includegraphics[width=\columnwidth]{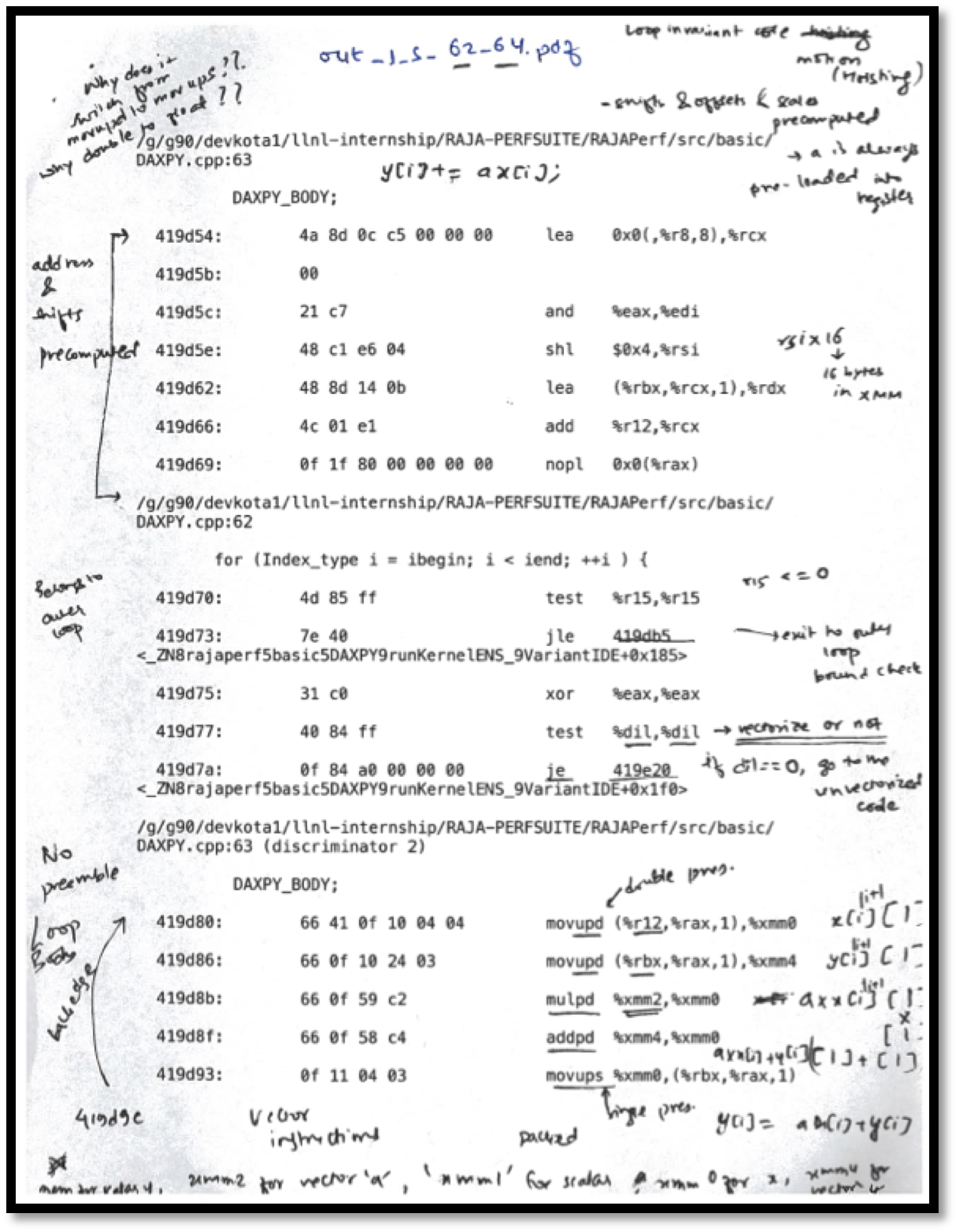}
 \caption{Annotations made on the disassembly of a benchmark program for vector addition during immersion study.}
 \label{fig:disassembly_artifact}
\end{figure}

When the domain experts initially trained the visualization experts in this process, they started with small enough examples that the recommended workflow was almost entirely pen and paper (\autoref{fig:disassembly_artifact}). The learner printed a filtered version of the disassembly. As they were able to correlate with source, they annotated the disassembly with variables and structures from source along with evidence of optimizations.

A complimentary approach is to use an integrated debugging tool which aids navigation between source code and disassembly, but is more focused on correctness debugging than optimization. We found most people we spoke with viewed files directly rather than through a debugger when trying to understand compiler optimizations.


\subsection{Related Work}
\label{sec:related_work}

Several tools link source and disassembly~\cite{totalview,intel-vtune,godbolt,ghidra,radare2,idapro,Roberts2005} for debugging or reverse engineering. Intel Vtune~\cite{intel-vtune} can incorporate profiling information---metrics about how fast the code ran. The Godbolt Compiler Explorer allows fast switching between compilers and options, linking across the multiple generated assembly files, and SeeSoft~\cite{Eick1992}-style file navigation. However, it does not scale to large programs. Reverse engineering tools~\cite{ghidra,radare2,idapro} also incorporate a visualization of the CFG, though with limited selection and filtering.

Other approaches prioritize either the source or disassembly. Rivet~\cite{Stolte1999Rivet} visualizes how instructions are scheduled on superscalar processors. Instructions within a window of time are linked back to source code. The focus is on the processor's scheduling, rather than the choice of instructions. PSE~\cite{Koppelman2014PSE} visualizes instructions collected while a program executes. It can therefore show performance metrics, but does not incorporate source code. Baum et al.~\cite{condcomp-baum-2019} present a visual tool for exploring conditionally compiled variants of programs. The focus of the tool is displaying what portions of the source code remain, rather than the resulting disassembly. 

Linking between source code and call graph has also been used in applications like performance analysis~\cite{Adhianto2010HPCToolkit} and software maintenance~\cite{Karrer2011Stacksplorer}. Several tree-metaphors have been used for call graphs including indentation~\cite{Adhianto2010HPCToolkit, Lin2010, Geimer2010Scalasca}, node-link diagrams~\cite{Krinke2004programslice, DeRose2007, Lin2010, VAtrees}, icicle timelines~\cite{Bezemer2015FlameGraph}, and sunbursts~\cite{Adamoli2010Trevis}. As the call graph served as an auxiliary view and following design study methodology guidelines of `satisfying rather than optimizing,' we use an indented tree and node-link view for different subgraphs of the call graph, leveraging familiarity of our users while supporting their tasks.

While many of these visualizations share core views and features with \projectname, we found no tool or design that suited the needs of our target audience in terms of other important elements such as annotation, filtering, scalability, and integration with structural views like the CFG. Furthermore, like the visual designs, the integrated analyses were for other purposes, not for exploring compiler optimizations. Despite the similar domains of these projects, the task differences led to a different design. We discuss related issues of transferability further in \autoref{sec:reflection}.

\section{Methodology, Data, and Task Analysis}
\label{sec:task_anls}

In conducting this design study, we followed the guidance of Sedlmair et al.~\cite{sedlmair-designstudies-2012}. We detail our collaboration below (\autoref{sec:methodology}) as well as the resulting data (\autoref{sec:data}) and task analyses (\autoref{sec:hier-task-anls}).  \autoref{sec:ccnav} then describes the resulting visual analytics approach.

\subsection{Design Process}
\label{sec:methodology}

Our team consists of two visualization experts, an HPC applications expert, an expert in (HPC) program analysis and tools, and a software developer. Two additional HPC experts attended the early project meetings as well. The applications expert represented the typical front-line analyst, though the program analysis expert also had goals in understanding compilation.

The program analysis expert approached the visualization experts upon seeing their prior work with visualizing CFGs~\cite{cfgexplorer-2018}. He wanted to extend the work to support his use case of optimizing compilations. 

The resulting collaboration has been ongoing for 18 months with video conferences scheduled every other week. These meetings included discussions of the available data, the analysis needs, and the development and deployment of \projectname, including both the visualization front end and the analysis software backing it. Copious notes were generated each meeting. Demonstrations via screen share were frequent, with the domain experts modeling their tasks using a combination of existing tools and the presented prototype as driven by the lead visualization expert in a pair analytics~\cite{hernandez-2011-pair,elmqvist2012-pair} fashion. 

A visualization expert (the lead author) also spent three months on-site with the domain experts. With their guidance, he performed their current workflow to better understand their tasks. We discuss the immersive elements of our collaboration further in \autoref{sec:reflection}.

\subsection{Data}
\label{sec:data}

The input data for \projectname is a compiled executable and its source code, the former of which can be disassembled into disassembly code. Both source and disassembly code are text data. There may be multiple source files associated with a single executable file.

Through a custom static analysis tool built using Dyninst~\cite{dyninst-2000} by the program analysis expert, we derive a mapping between lines of source code and address ranges in disassembly code. Note that this mapping can be many-to-many. We also derive a control flow graph, loops within that graph, a mapping between source code variables and disassembly, and annotations regarding disassembly addresses of inlined functions.

There are limitations to the automated analysis. For example, different compilers report varying amounts of information, which affects the completeness of the mapping between source and disassembly. The program can't match some variables with registers in the disassembly. Some function names, which are mangled into unique identifiers by the compiler, do not get properly de-mangled. Experts must combine the automated assistance with their awareness of compiler reporting limitations and knowledge of the domain.

The programs of interest to our collaborators are sizable and complicated, using many advanced features and libraries. Thus, we can make few assumptions. For example, in one case we found a de-mangled function name that was (correctly) 137,777 characters long.


\subsection{Task Analysis}
\label{sec:hier-task-anls}

The ultimate goal of our collaborators is to determine a combination of source code changes, compiler choice, and compiler flags that will achieve improved performance. The domain experts are aware of strategies the compiler can take, so they analyze the results of the compilation to determine where there is room for improvement.

Following the ethos of understanding tasks in the context of high level goals~\cite{hiertask-salmon-2010,typology-brehmer-2013,bridgegoals-lam2018,idmvis-2019}, two visualization experts independently coded the observation notes and then met to discuss and ultimately generate the task hierarchy. We found no task typology mapped well to the low-level operations, which were frequently correlating concepts (e.g., source code lines to disassembly lines) and identifying known structures. We present the higher levels below and summarize the lower levels in text. The full hierarchy is in the supplemental materials.

\begin{figure*}[tb]
 \centering 
 \includegraphics[width=\textwidth]{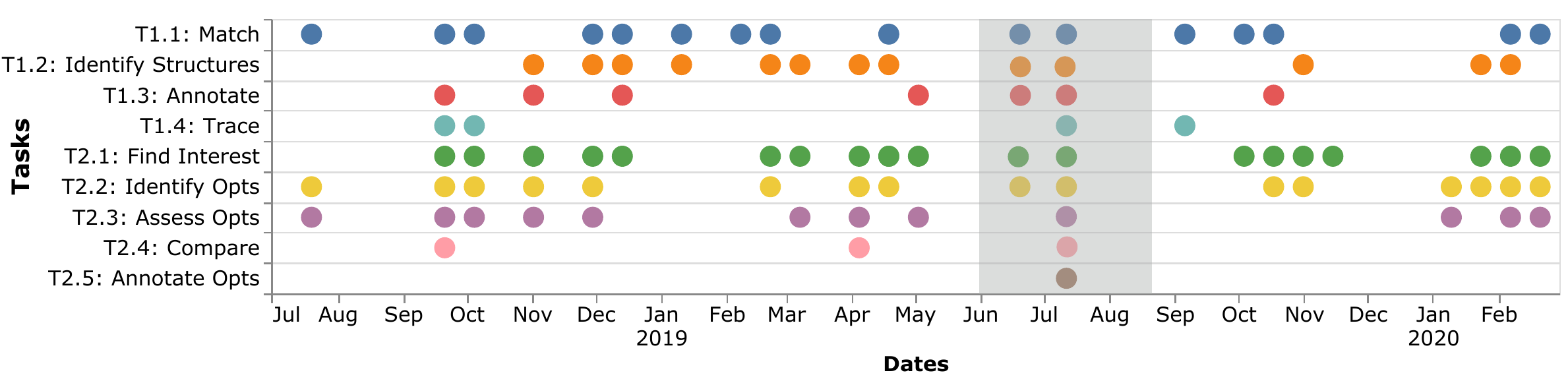}
 \caption{Tasks appearing in our meetings and on-site immersive study. The on-site period is highlighted with a gray background. Tasks related to understanding the disassembly and finding areas of interest for optimization dominated. These tasks are necessary to perform the less prevalent tasks but are difficult in their own right, thus our design study focused on them.}
 \label{fig:task_timeline}
\end{figure*}

Focus on particular optimizations or analysis strategies varied from meeting to meeting, though the overall goal did not change. Similar to Williams et al.~\cite{williams2020-moving}, we used the persistence of tasks over time to prioritize the design and implementation of \projectname features. \autoref{fig:task_timeline} shows when tasks were demonstrated or discussed in our interactions with domain experts.

\begin{myEnumerate}[label=]
\item Goal: Understand performance / Identify optimizations 
\label{Goal}
  
  \begin{myEnumerate}[label=T\arabic*]
      \item Understand/Identify compiled structure
        \label{T1-understand}
    \begin{myEnumerate}[label*=.\arabic*]
        \item Match source code with binary code
        \label{T1s-match}
    
        \item Identify/Relate structures with code
        \label{T1s-structures}
        
        \item Annotate relations
        \label{T1s-annotate}

        \item Trace variable
        \label{T1s-trace}

    \end{myEnumerate}
    
    \item Understand optimizations
    \label{T2-optimization}
    \begin{myEnumerate}[label*=.\arabic*]
            
        \item Find areas of interest
        \label{T2s-interest}
        
            
 
            
        \item Identify optimizations
        \label{T2s-identify}
        
        \item Assess optimizations
        \label{T2s-assess}
        
        \item Compare generated code
        \label{T2s-compare}
        
        \item Annotate optimizations
        \label{T2s-annotate}
        
        \end{myEnumerate}
    \end{myEnumerate}
\end{myEnumerate}

We found two major groupings of tasks: understanding and interpreting the disassembly itself (\ref{T1-understand}) and understanding what optimizations were applied in it (\ref{T2-optimization}). When we started this project, we expected the focus would be on \ref{T2-optimization}, specifically comparisons across parameters (\ref{T2s-compare}). However, our initial collaborative analysis sessions showed us that simply understanding how what we were looking at related to source (\ref{T1-understand}) was a significant hurdle.

\vspace{1ex}

\noindent\textbf{T1: Understand/Identify Compiled Structure.} The disassembly represents what the compiler did. To understand what the compiler did, analysts must match the disassembly and the source (\ref{T1s-match}). Typical queries are ``What disassembly matches these lines of code?'' or ``What are these lines of disassembly doing with respect to the source code?'' As code structures like functions and loops both help organize the code, identifying those structures in particular are a common task (\ref{T1s-structures}). Once these first sub-tasks are done, the disassembly may be annotated (\ref{T1s-annotate}), e.g., marking a register by its associated source code variable or marking an address range with a line of code, loop, or function. Another way to understand disassembly in the context of source is to trace (\ref{T1s-trace}) a source variable through the disassembly.

\vspace{1ex}

\noindent\textbf{T2: Understand Optimizations.} Analyzing how well a compiler has optimized some code is typically focused on the instructions that will be run the most. Thus, the first sub-task would be finding those areas of interest (\ref{T2s-interest}). This is often a winnowing task---decreasing the data to a specific function, loop, or even line of code. However, it may also be a search task, like identifying anomalous code performing an unreasonable number of operations. Thus, the entirety of the code must be accessible.

Once the area of interest has been found, the analyst will try to identify the optimizations present (\ref{T2s-identify}), such as inlining, vectorization, code variants, or unrolling, and make an assessment (\ref{T2s-assess}) regarding whether the optimizations applied are appropriate or if any are missing. Performance metrics, if available, can also be used to assess the efficacy of the optimizations. Typical queries include ``Is this loop vectorized? What about its nested children?'' and ``How much inlining is there?'' As with \ref{T1-understand}, discoveries are annotated (\ref{T2s-annotate}) during the analysis process.

The identification of the absence or presence of possible optimizations and their effect on performance may be further supported by comparing disassembly generated with different source, using different compilers, or using different compiler optimizations (\ref{T2s-compare}). However, this operation is limited by the difficulty of understanding even one compilation.

\subsection{People}
\label{sec:people}

The target audience of our project is application or program analysis experts with experience reviewing disassembly code. We focus on those who are interested in optimization, but there is overlap with those who are trying to debug compilation or build issues as well. These expert users are familiar with DWARF and other debug data, as well as the limitations in collecting and reporting it.

\section{\projectname: Compilation Navigator}
\label{sec:ccnav}

The existing workflow of our collaborators involved understanding the compiled
code using multiple tools to create or view different artifacts related to the
program such as the source code, disassembly, and debug information. The
process of relating between these artifacts requires a large amount of context
switching between the different programs and is both labor-intensive and
time-consuming. 

Through our regular meetings, we iterated on the design of \projectname,
discovering in addition to the findings of our data and task analyses
(\autoref{sec:task_anls}), that experts in this style of program analysis: 1)
have many strategies in navigating code artifacts, indicating a highly linked
multi-view system could streamline their strategies, and 2) they generate new data
and new data arrangements in the form of supporting annotations and graphs and
that some of this generation could be automated. Thus, we developed a custom
analysis program designed in tandem with a highly-coordinated multi-view
system to better serve the needs of compilation analysis.

We balanced the effort in our design by focusing more on the tasks most numerous and persistent across time (\autoref{fig:task_timeline}). These were the tasks necessary to perform the other tasks: those related to understanding the disassembly and finding areas of interest for optimization.

The input to \projectname is a binary file compiled with debug information. We
derive the rest of the data through a custom analysis program developed for
this project (\autoref{sec:optparser}). We first describe the
views and interactions (\autoref{sec:views}) which are based in our task
analysis and observations and drove the development of the automated analysis.

\subsection{Views}
\label{sec:views}

\projectname is composed of multiple views which can be arranged, resized,
closed, and re-opened by the user via a flexible window management system. We
describe these views and their relation to our tasks.

\vspace{1ex}\noindent \textbf{Source Code View.} (\autoref{fig:teaser}(a)) The source code view
displays a single source code file. By default, it displays the one with the most data, but the file can be changed in the interface. Multiple lines can be selected
and will be highlighted across other views, supporting the task of matching
the source code and disassembly (\ref{T1s-match}). Lines with no mapping are grayed out. We chose not to use syntax highlight to conserve the use of color and because our domain experts did not consider it a priority. 

\vspace{1ex}\noindent
\textbf{Disassembly View.} (\autoref{fig:teaser}(b)) The disassembly represents the ground truth of
the compiled program. One strategy commonly employed by users was to use
linked navigation to get close to an area of interest not otherwise
selectable with information from our automated analysis and then search by
scrolling from there, so we include it in its entirety. This view also supports
multi-line linked highlighting in support of
\ref{T1s-match}. 

When available, we modify the instruction text to include the associated source
code variable name. We denote this by striking through the register name
and presenting the source code with a pink background. This feature supports our
annotation (\autoref{T1s-annotate}), structure identification
(\ref{T1s-structures}), and variable tracing (\ref{T1s-trace}) tasks.

\begin{figure}[bt]
 \centering 
 \includegraphics[width=\columnwidth]{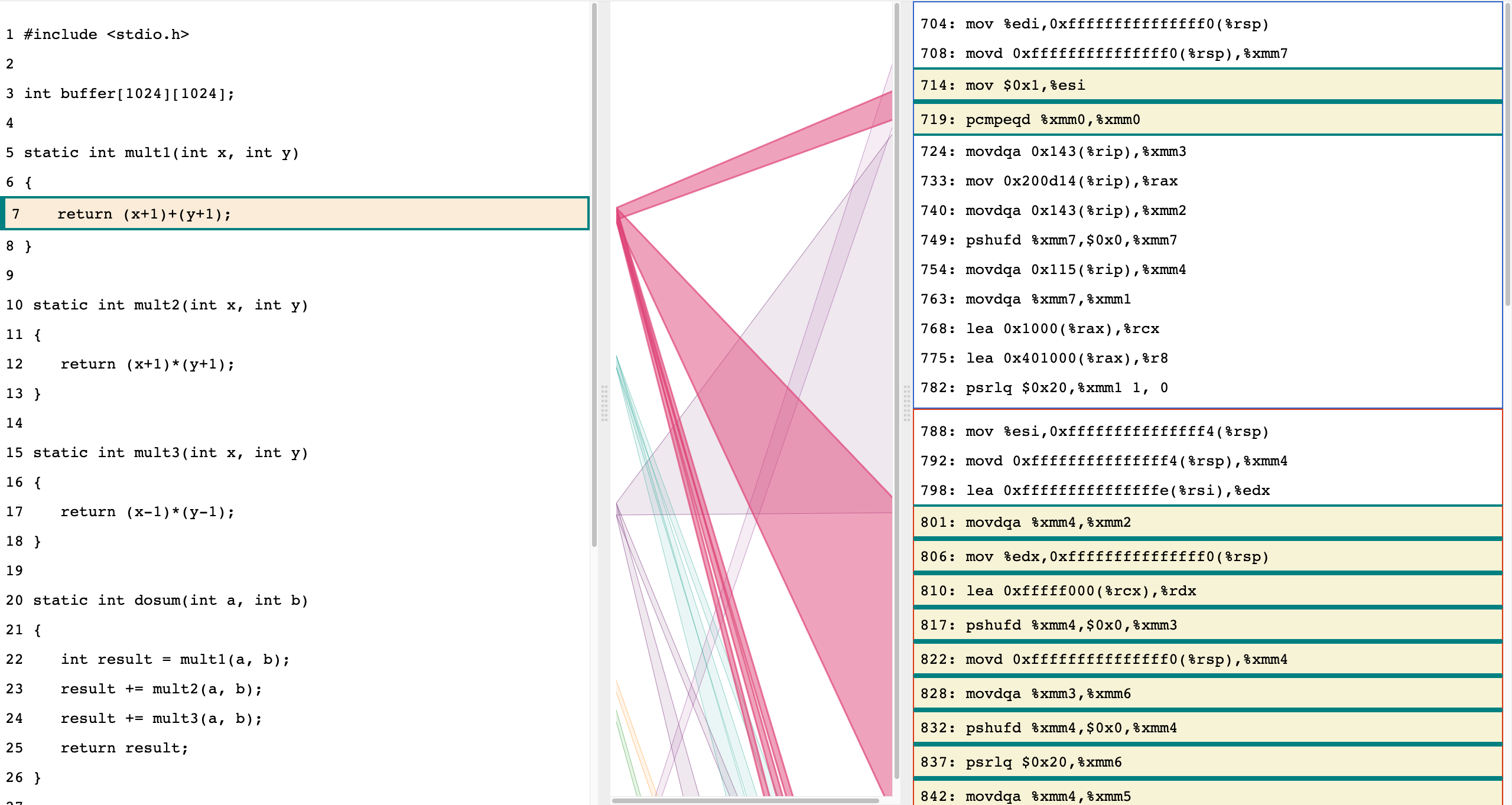}
 \caption{An early prototype visualizing the mapping between source code and disassembly. Even in this toy example, a single line of code is dispersed across the disassembly. 
 Early prototypes like this one also clarified the importance of the CFG in conveying structure.}
 \label{fig:dispersion}
\end{figure}

\vspace{1ex}\noindent
\textbf{Control Flow Graph (CFG) View} (\autoref{fig:teaser}(c)) The CFG view
shows a subgraph of the full binary CFG, based on the current selection. Prior work on CFGs by the visualization experts led to this project. However, early meetings indicated matching of source and disassembly was the main workflow. Thus, our initial prototypes did not include a CFG. (See \autoref{fig:dispersion} for one such example prototype.) In subsequent meetings, we observed our domain experts had difficulty understanding structures such as loops (\ref{T1s-structures}) with only matching or nesting. We thus chose to provide such context with a CFG view.

We chose the visual design and layout from CFGExplorer~\cite{cfgexplorer-2018}
as (1) that design was the impetus for our collaboration and (2) the tasks fulfilled by a CFG in this project matched well with the tasks in CFGExplorer. The design is a node-link diagram with a modified Sugiyama layout~\cite{sugiyama-1981} which prioritizes
drawing loops similar to by-hand diagrams of small CFGs, matching the mental
model of compilation experts. The convex hull of loop nodes are drawn with an
orange background, with nested loops having a darker shade of orange. 

Instead of showing all contained instructions as CFGExplorer did, we
show the block ID and its containing function. We found including all
instructions led to very long nodes which obscured the graph topology and worked against  \ref{T1s-structures}. After making the design choice, during one of our pair analytics sessions, our collaborators
commented they could see the global structure and the connected components in
the graph. They noticed that a program we were viewing had a
disconnected CFG, leading to the insight that library and initialization code were present but unable to be retrieved by the automated analysis.

Another change from CFGExplorer is filtering the graph to a $k$-hop region of
interest around selected basic blocks. Our data creates CFGs that are too
large for Sugiyama-style layouts. To support the winnowing of data to find
areas of interest (\ref{T2s-interest}), $k$ is configurable via the
interface, with a default of $k = 3$ determined through our users' experience.

Basic blocks (nodes) in the CFG can be selected individually or by brush and
will update all views (\ref{T1s-match}). The CFG view also supports
panning and zooming. 

\vspace{1ex}\noindent
\textbf{Highlighted Items View} (\autoref{fig:teaser}(g)) The highlighted
items view lists the highlighted source lines, disassembly lines, and
basic blocks without context. As highlighted items are often dispersed
across large ranges of source lines (see \autoref{fig:dispersion}), this view provides a way to examine them
together when the content is more important than the context, e.g., when
assessing the use of instruction types (\ref{T2s-identify},
\ref{T2s-assess}) or the presence of variables (\ref{T1s-structures}).

\vspace{1ex}\noindent \textbf{Function Inlining View}
(\autoref{fig:teaser}(f)) Function inlining is one of the most common
optimizations performed by compiler and is of great interest to our
collaborators. Thus, we design a separate panel for inlining information to
help identify (\ref{T2s-identify}) and navigate (\ref{T2s-interest}) to them. 

We use a selectable, collapsible indented tree with non-inlined functions as the top level and only in-lined children beneath them. Selections in other views will filter this one. 

The function inlining hierarchy can get quite large as functions may be
inlined in multiple places and inlining chains into libraries or kernel code can be tens of layers deep. We include an autocomplete search feature to further support navigation (\ref{T2s-interest}) and a reset control to restore the full view.

Function calls and therefore inlining forms a hierarchy, so a tree visualization is intuitive. We picked the collapsible indented tree to preserve readability of function names and efficiently use screen space given their size and deep nesting. Also, while a direct inlining view is not common, our audience is familiar with collapsible indented trees for navigating call stacks or file systems.

\vspace{1ex}\noindent 
\textbf{Loop Hierarchy View} (\autoref{fig:teaser}(e)) Identifying (\ref{T1s-structures}) or navigating to (\ref{T2s-interest}) a particular loop is a common operation, so we chose to directly support it by creating a loop-centric view. Consistent with the function inlining view, selections in other views will filter this one, providing loop context to those other views. 

We designed this view to show the nesting hierarchy of loops as a collapsible indented tree with linked selection to the other views. The top level is the containing function, matching the policy of the function inlining view.

There is no standard way to name loops, nor can the appropriate line of source code be derived with suitable consistency. Thus, we assign multi-part IDs to indicate nesting behavior and rely on the analyst to interpret them further using the other features of \projectname.

Explicitly showing the loop hierarchy is not common and thus there is no standard view. Our rationale for using a collapsible indented tree is similar to that of the function inlining view and as it was added second, we chose to keep the designs as consistent as possible.

\vspace{1ex}\noindent
\textbf{Call Graph View} (\autoref{fig:teaser}(d)) The call graph view shows a subgraph of the full call graph, with all functions reported by our analysis regardless of whether they were inlined. This view provides a way to relate selected disassembly to the functions and call stack in support of \ref{T1s-structures} and \ref{T1s-trace}. Inlined calls are shown with a dashed red line to help identify them (\ref{T2s-identify}). We chose a node-link diagram to display the call graph as analysts performed navigation
tasks~\cite{lee-2006,ghoniem-2004,keller-2006} on the graph. This view supports linked selection with the other views.

\vspace{1ex}\noindent
\textbf{Variable Annotation View} (\autoref{fig:teaser}(h)) Annotating the disassembly with source code variable names is a common task (\ref{T1s-annotate}). While our automated analysis provides a best-effort annotation, it is incomplete. We allow the user to manually add annotations with this view. The view further summarizes all active renamings.

\subsection{Design of Linked Highlighting and Navigation}
\label{sec:linking}

We follow a consistent scheme for selections across all the panels. On performing any selection, the corresponding items in all text and node-link panels are highlighted with teal border. To reduce scrolling, the views automatically scroll to the first highlighted item in text views and center on the first highlighted node in node-link views. We do not alter the zoom for node-link diagrams as users found it disorienting. 

For our indented tree views, we filter to matching items rather than highlighting them. The ordering of top-level nodes in these trees are not consistently related to source code structure. The ordering instead is an artifact of the analysis, and thus the context is less meaningful, so we filter these hierarchy views more aggressively.

We support a consistent interaction mechanism across all the views where `click' interactions select single items (e.g., line of code, node). Text views support range selection through `click and drag', while node-link views support it through brushing.

The one exception to our linking is the Highlighted Items View. We found linking this view resulted in mis-clicks and mis-selections as people focused on this window. However, based on our Evaluation (\autoref{sec:evaluation}), we are considering changing this policy in the next iteration.

All linking and filtering is calculated based on disassembly address. Unit items in a view (e.g., line of source code, instruction address, basic block, loop, function) can be represented as a list of corresponding address ranges from our automated analysis. Thus, any selection is translated into a list of (non-contiguous) address ranges which is then used to query matching items in all the other views.

We use interval trees to speed up queries. Specifically, we create 4 interval trees for storing the address ranges associated with i) lines of source code,
ii) basic blocks, iii) functions, and iv) loops. These trees are also used to reconstruct the inlining tree and loop hierarchy.

\subsection{Automated Analysis}
\label{sec:optparser}

\projectname takes as input a binary file compiled with debug information and
from it retrieves or generates all additional data used in the visualization.
This includes: disassembling the binary into disassembly code, retrieving the
source code (if reachable), reporting available mappings between source and
assembly (including variable names), reporting inlined functions, detecting
loops, and generating a control flow graph and call graph. These features were
added iteratively, matching with the visualization design.

The data generated by the automated analysis is incomplete by nature due to
limitations in what an individual compiler will report and limitations in
state-of-the-art static program analysis. For example, most lines of code do
not map to disassembly. These limitations are one reason a completely automated solution is infeasible, leading to our design a visual analytics system to combine partially automated analysis with expert knowledge and intuition. Furthermore, the limitations drive our design to support multiple workflow to target disassembly, as any common workflow may fail in a particular situation.

The automated analysis also provides the front end visualization with shortened names of strings greater than 256 characters by eliding the middle characters in long function names. Our domain experts indicated further elision was too much.

\subsection{Implementation} \projectname is a browser-based client-server
application. The automated analysis is written in C++ using the Dyninst
API~\cite{dyninst-2000}. The server returns the output in JSON format
in a RESTful manner. The client is written in Javascript using
D3.js~\cite{bostock-d3-2011} with Dagre~\cite{dagre} as the base layout for directed graphs. Flexible window management is implemented using
GoldenLayout~\cite{goldenlayout}. The autocomplete search is supported by the awesomplete library~\cite{awesomplete}. We use the flatten-js interval tree library~\cite{flattenjs} to speed up the calculation of addresses across our linked views.

\section{Evaluation}
\label{sec:evaluation}
To evaluate the effectiveness of \projectname, we conducted evaluation sessions with four participants. 

\subsection{Evaluation Session Design}
\label{sec:eval-sessions}

The evaluation sessions were 90 minutes in length and consisted of an initial briefing, an overview and demonstration of \projectname, tasks for the participants, a semi-structured interview, and a debriefing. The overview and demonstration used a small example dataset. With questions, the demonstration portion was approximately 25 minutes long. All evaluations were conducted remotely over video conference, with the facilitator conducting the evaluation sharing his screen.

\vspace{1ex}

\noindent\textbf{Participants.} There were four participants. The first, P0, was a graduate student with experience with disassembly, but not performance analysis. The other three, P1-P3 were professionals who often perform compilation performance analysis. P3 attended design meetings for the first two months of the project, but had not seen any prototypes in the intervening 16 months. P0-P2 were not involved in the design. P2 had a time constraint, so their session was limited to 70 minutes.

\begin{figure*}[bt]
 \centering 
 \includegraphics[width=\textwidth]{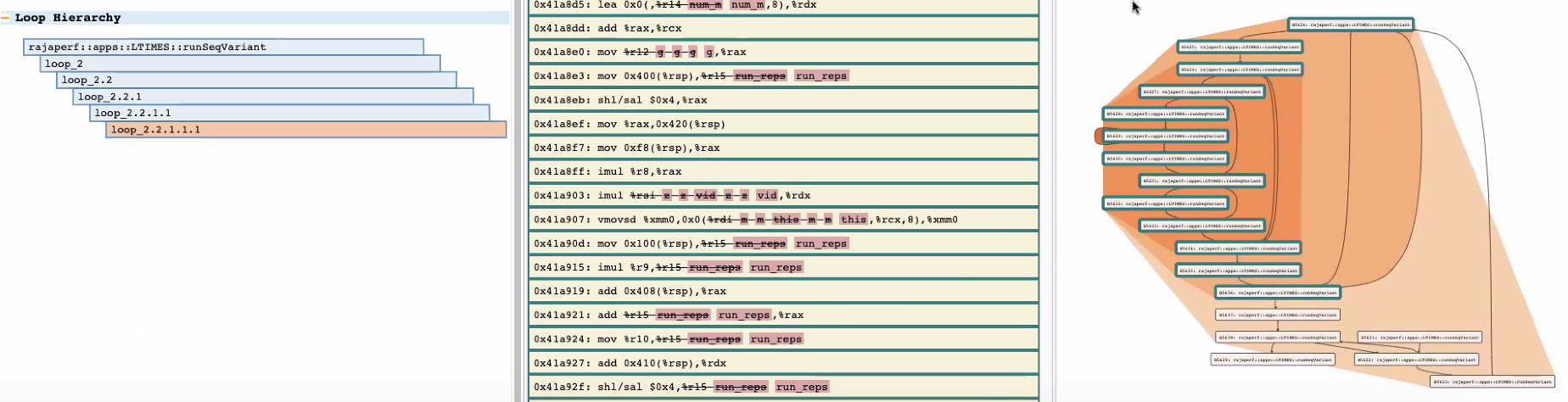}
 \caption{Drilling down into the loop hierarchy (left) reveals nested loops in the CFG subgraph (right). Associated disassembly (middle) is highlighted. Several registers have been automatically annotated with variable names from source code.}
 \label{fig:rajadrilldown}
\end{figure*}

\vspace{1ex}

\noindent\textbf{Pair Analytics.} We employed pair analytics~\cite{hernandez-2011-pair,elmqvist2012-pair} in our evaluation. Following this method, we encouraged participants to provide specific instructions (e.g., `click', `scroll', `go to the CFG view') to the facilitator who would then perform them. Participants were also encouraged to `think aloud' as they directed these actions. The facilitator answered questions from the participants. One author acted as facilitator while another took notes. 

Our choice of pair analytics was driven by our goal of evaluating whether the design of \projectname supported {\em analysis workflows}. We wanted the focus to remain on the analysis rather than the troubleshooting associated with using an in-development system or the learning curve of a new complex system. We also value the benefits of pair analytics in encouraging participants to communicate their actions and thoughts. However, we recognize there are potential biases associated with pair analytics, which we discuss further in \autoref{sec:threats}.

\vspace{1ex}

\noindent\textbf{Evaluation Dataset.} For the task part of our session, we used the LTIMES application of RAJAPerf~\cite{rajaperf}. RAJAPerf is a proxy application for assessing performance and portability of HPC code. The core computation of LTIMES is a quadrupally nested loop, of which several versions are implemented in the same C++ file. We compiled LTIMES using Intel C++ Compiler v19.1.0~\cite{icpc} with flags \texttt{-O3} and \texttt{-g}. P1 and P3 had worked with RAJAPerf before and P2 was familiar with it, though none were particularly familiar with LTIMES.

\vspace{1ex}

\noindent\textbf{Tasks.} Participants were asked to perform evaluation tasks of increasing complexity. Our initial task list included basic tasks like identifying what was inlined in a line of source code. After sessions with P0 and P1, we determined these tasks were too elementary and omitted them to afford more time to the more open-ended tasks. We list the tasks given to all participants below with their corresponding task abstraction items from \autoref{sec:hier-task-anls}:

\begin{enumerate}[label=E\arabic*.]
    \itemsep=0.25ex
    \item Identify the assembly of a loop containing a selected line of source code (\ref{T1s-match}, \ref{T1s-structures}, \ref{T2s-interest})
    \item Identify/Assess vectorization in that loop (\ref{T2s-identify}, \ref{T2s-assess})
    \item Compare/Assess multiple variants in the source code (\ref{T2s-compare}, \ref{T2s-assess})
\end{enumerate}

\subsection{Evaluation Task Results}

We summarize our observations of task performance below. A more detailed account is available in the supplemental materials.

\vspace{1ex}\noindent \textbf{E1: Identify the assembly of a loop containing a selected line of source code.} Because a loop spans multiple lines and the mapping between source code and disassembly is imperfect, this task requires more than straightforward highlighting. All participants were able to complete this tasks with different strategies:

All participants started by asking to click on the first line of the loop in the source code, which highlights the directly corresponding disassembly, but not the entire loop. They all recognized this fact. P0, P1, and P2 next examined the loop hierarchy view. P0 and P1 asked to click on the loop hierarchy view to highlight the assembly, while P2 returned to the source code and asked for a range selection. Both strategies result in the targeted selection. P2 followed up by asking to click on the loop hierarchy view, verifying the selection was the same.

P3 instead looked at the selected items view, finding the loop index variable `z' annotated and were satisfied they had found the code. When asked for the loop name in the loop hierarchy, they asked to click on the top level loop \texttt{loop3}, similar to the other participants. Observing that both source code and loop hierarchy have five levels of nested loops, P3 guessed the correct loop.

\vspace{1ex}\noindent \textbf{E2: Identify/Assess vectorization in that loop.} P1, P2, and P3 said they would look for vector instruction, but noted they did not recall or know them by name. P0 required some background knowledge on vectorization. The facilitator instructed that the presence of a vector register would indicate vectorization. P0, P1, and P2 were suggested names of vector registers.  

P0, P1, and P2 started by asking to click on loop 3.1 in the hierarchy view while P3 asked to click on the body of the innermost loop in the source code, explaining they wanted to look for arithmetic instructions and unrolling. All participants then went to the selected items view. P0 and P3 asked to scroll through them while P1 and P2 chose to search (\texttt{ctrl-f}). They all discovered vector registers and instructions. P1, P2 and P3 concluded the loop was vectorized. P0 followed up by returning to the disassembly view and asking to select the found instruction there. They verified the loop was selected in the source code and only then asserted the loop was vectorized.


\vspace{1ex}\noindent \textbf{E3: Compare/Assess multiple variants in the source code.}  The LTIMES application has several versions of the same computation. In this task, we focused on two: a) a ``base-sequential'' (``Base'') version with nested four loops, and b) a ``RAJA-sequential'' (``RAJA'') version where loops are implemented using RAJA constructs and thus the quadruple nesting is not explicitly written in the source file. Some participants also chose to look at a third variant, ``lambda-sequential'' (``Lambda'') which is like Base, but uses a lambda function for the body. 

The task was free-form by design. Each participant approached it with a different strategy. P1, P2, and P3 were able to draw conclusions. P0 was able to isolate the RAJA disassembly, but said they did not know how to assess differences due to lack of experience in such analyses.
 
Identifying each variant's disassembly and assessing the optimizations were key sub-tasks. As in E1 and E2, they typically started by selecting the source code or the loop hierarchy, switching between these views to further their search while using the other views to examine the changes in the selection. 

Selecting the RAJA disassembly was the most tricky, because while it could be selected from the full loop hierarchy (P0, P1), selecting the source code (P0-P3), retrieved only a few instructions and filtered out all loops. All participants recognized this limitation of the automatic matching with the source and found the full target disassembly either by searching for RAJA code in the inlining view (P1) or finding related lines of source code (P0, P2, P3). From there, P0, P1, and P2 used the loop hierarchy to further navigate the disassembly.

P3 did not recall that elements in the loop hierarchy could be clicked to drill down and instead examined the selected items view. Spotting the annotations in the disassembly for variable \texttt{phidat}, P3 hypothesized they were looking at the data setup, but wanted arithmetic instructions that would indicate the loop body. They switched to the full disassembly view and found some non-highlighted arithmetic instructions and said ``that's completely what we want to see.''

While navigating the code, the participants all considered the CFG View. However, in many cases they noted it was not enough information because it showed function names and not instructions (P1), often returned disconnected nodes due to filtering (P2), or was too low level and lacked context (P3). P2 used the call graph view to reason why the nodes were disconnected in their selection. P1 identified the quadruple-nested RAJA loop in the CFG View (\autoref{fig:rajadrilldown}), and from there identified candidates for the preamble and postamble loop instructions.

In assessing variant similarities, P1 noted the code structure was similar between RAJA and Base, but RAJA was obfuscated by a long call stack. P2 and P3 remarked both versions had similar vectorization. After drilling down further in the loop hierarchy, P2 hypothesized that both versions have everything inlined, but there is more overhead in the RAJA version due to the indirect call. This is consistent with performance data not used in the evaluation.

\begin{figure}[bt]
 \centering 
 \includegraphics[width=\columnwidth]{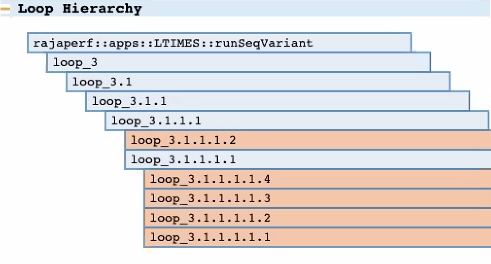}
 \caption{Loop hierarchy view. Evaluation participant P1 determines the leaves are four variants of the same loop, generated by the compiler to aid loop unrolling.}
 \label{fig:loopvariants}
\end{figure}

P1 also compared the Base and Lambda versions, finding them to be similar. By navigating down the loop hierarchy, they came to the conclusion that the inner loops (\autoref{fig:loopvariants}) in both versions were vectorized and that the leaf loops are ``fixing up the ends for the vectorization unroll.'' They repeated the process with the Base version, confirming their hypothesis.

\subsection{Semi-structured interviews}
Participants were asked what tasks were easy or difficult and what features they would like to see. We summarize the resulting discussions.

Participants generally liked the linking between all views (P0, P2, P3), with some remarking that the variable renaming is helpful in decreasing the need to switch between multiple sources (P0, P3). P3 said of the linking, ``It already beats pawing around in something like VTune'' and ``I gotta say that variable renaming thing changes so much in trying to navigate this thing.''

Participants also remarked other views were useful for overview and navigation, including the loop hierarchy view (P1, P2), the function inlining view (P2), and the CFG and call graph views (P0). P1 noted the CFG picked up loops well in the RAJA version, but not the base version. P3 found the selected items view convenient.

Participants expressed difficulty with the drill down behavior in the loop hierarchy (P2, P3). P1 noted the autocentering of the source code was disorienting and wanted more text in the CFG nodes. Suggestions for new features included a back button and history (P2, P3), annotations of loop preamble, postamble, and body (P1), keyboard shortcuts (P2), and pop-out windows for more space (P3).

P2 summarized their remarks with ``I'm kind of excited to try this out on a couple of different things.'' They later added in email a situation where they previously compared different compilation flags for three versions of the same source code. They manually created a rough equivalent of the selected items view and produced a diff of the results. They remarked \projectname would have been ``easier, faster, and cleaner'' if it supported this kind of comparison. 

P3 shared that he has compared program performance across compilers, noting he would do exactly the tasks from the evaluation session when trying to determine if the compiler applied the changes correctly.

\subsection{Evaluation Findings}
\label{sec:findings}

The participants completed all of the tasks with the exception of the non-expert P0 on the comparison task. Participants employed a variety of strategies in each task. We consider this to be positive evidence of the system's flexibility in supporting compilation analysis. Compilation analysis is complicated and often requires clever ways to probe. 

For example, in E3 the RAJA disassembly proved non-trivial to isolate. Participants used multiple views in sequence for selection (source, loop hierarchy, function inlining) and multiple views to assess the results (disassembly, selected items, loop hierarchy, CFG, call graph). This meandering style of navigation, where participants are free to consider different facets, matches our task analysis observations. Multiple strategies can further allow analysts to verify discoveries, as we saw P0 do in E1. 

Participants also expressed positivity regarding linked navigation, but  noted a lack of tool-maintained history supporting their exploration. We observed some participants repeat actions to return to prior views, further underscoring this potential area of improvement.

Another goal of \projectname was to aid users with their mental model. We observed all participants using the nested nature of the loop hierarchy to navigate. P1 was able to match disassembly instructions with higher level loop constructs using the CFG view.

Through the evaluation tasks, the participants performed tasks from our task abstraction. Source-disassembly matching (\ref{T1s-match}), loop identification (\ref{T1s-structures}), and finding areas of interest (\ref{T2s-interest}) were sub-tasks in all evaluation tasks. Participants identified (\ref{T2s-identify}) and assessed optimizations (\ref{T2s-assess}) in E2 and E3. Participant P3 used annotations (\ref{T1s-annotate}) in E3. We interpret this as validation of our task analysis and of \projectname's ability to support those tasks.

Though comparison is not supported explicitly, P1-P3 were able to compare (\ref{T2s-compare}) results of different versions of the same code in E3. The only tasks not demonstrated were tracing a variable (\ref{T1s-trace}) and annotating optimizations (\ref{T2s-annotate}). These weren't required by the evaluation tasks and as they were the least performed tasks over our design study meetings, they were the lowest priority in our design.

All views were used by at least one participant to achieve some insight during the evaluation. We interpret this as validation for our choice of views. However, there was also some confusion caused by some of these views, many of which are related to selection and filtering choices, explained below. Another issue is the call graph view can get very wide---a more compact layout will require further research.

 Though the participants acknowledged limitations in debug information, these limitations still led to confusion regarding some of the selections. For example, participants clicked on the \texttt{for} loop line rather than range-selecting the whole loop. There was similar confusion with how much context was shown in the loop hierarchy and CFG. We believe both can be improved by showing more nesting context. We have since revised our CFG view to pull in the entirety of loops overlapping the selection rather than only those within the $k$-hop radius.

\subsection{Threats to Validity}
\label{sec:threats}

All participants came from the authors' institutions. In briefing they were told the purpose was to evaluate \projectname and determine issues for future iterations. However they may still have been inclined to give positive feedback.

The small participant pool in this evaluation limits its generalizability. Though the group was small, they demonstrated similar patterns in selecting disassembly of interest and using the source, disassembly, loop hierarchy, and selected items views. However, use of the inlining, CFG, and call graph views was more unique and should be interpreted as preliminary and with caution.

The remote nature of our evaluation required some concessions. All participants required a larger font size, decreasing screen real estate. Also, they could not point to anything on screen or ``take the reins,'' which may have changed their behavior.

All participants asked for reminders regarding details of particular views or interactions. Due to the complexity of both the visualization and their tasks, the demonstration was insufficient. Furthermore, the basic tasks performed by P0 and P1 (see supplementary materials) may have had a tutorial effect, accounting for some participant differences. 

While pair analytics may have alleviated some of the training issues, it may have also altered participant actions. For example, P3 did not recall they could click on the loop hierarchy and was unable to rediscover the functionality through remote pair analytics. We did not suggest it to them because they did not explicitly state that was their intended effect and thus we did not want to bias them.

In addition to limiting participant discovery, there is a complementary threat of leading, over-interpreting, or otherwise biasing participant actions. To mitigate biased findings, we explicitly recorded and reported where the facilitator made suggestions or answered complex questions. These are available in our more detailed description of participant actions in the supplemental material.

On reflection, in future projects with similarly complex tasks, we could combine pair analytics sessions for one set of participants with traditional sessions with another, thereby covering the limitations and enjoying the benefits of both. However, it may be difficult to recruit enough qualified participants.

\section{Reflections and Lessons Learned}
\label{sec:reflection}

We reflect on our design study and what we learned regarding transferability between design studies and immersion in the design process.

\vspace{1ex}

\noindent{\bf Transferability from a previous, highly-related design study was beneficial, but more limited than expected.}
One key outcome of design studies is transferable design knowledge, but it can be difficult to assess in what ways and to what extent such knowledge is transferable. This design study started in response to the domain experts seeing the visualization experts' previous work, CFGExplorer~\cite{cfgexplorer-2018}. The domain experts were particularly interested in the custom node-link layout of CFGExplorer and its linking to the assembly code. They wanted to directly {\em extend} CFGExplorer for their problem. The team thus began the project assuming previous work would be highly transferable and the process would be like a design iteration. However, in practice, we found the process more similar to a new design.

While two of the main data types (CFG and disassembly code) were the same between \projectname and CFGExplorer, the goals of our users, and thus the tasks the visualizations had to support, differed enough that we started the design anew. In CFGExplorer, domain experts are trying to recover parallelizable loops from the disassembly and CFG only. In \projectname, domain experts are trying to understand what optimizations were performed on their source. This shift in goals prioritizes source code in \projectname, a data type that was not available in CFGExplorer.

Despite our initial assumptions, we avoided premature design commitment to CFGExplorer by restarting our task analysis, questioning design choices frequently, and creating revolutionary prototypes. These correspond to the discover, design, and implement phases of design methodology. We did not assume any tasks going into our first meeting. The workflow described in that meeting emphasized the correspondence between source code and disassembly. We thus questioned the assumption to include the CFG and ultimately decided to omit it from our first design/prototype based on the experts' described operations. However, when we tried analyzing a problem using this prototype with the experts on our team, the value of the CFG became clear. The experts struggled to understand and recall how disassembly instructions related to loops, despite source code linking. This discovery led us to add the CFG view. Following this early design discovery, we continued to question our designs as the project evolved.

The custom node-link layout from CFGExplorer transferred because the primary tasks it served remained the same, albeit in a lesser role. In both CFGExplorer and \projectname, the node-link view serves in building a mental model from disassembly code and identifying loops. The two projects differ in their use of this view only in the level of detail required. In \projectname, some of the lower level operations, such as determining the loop bounds, were better served by the linked source code that was unavailable to CFGExplorer.  

\vspace{1ex}

\noindent
{\bf Immersive data analysis and prototyping activities had the most influence on our design.} Immersive activities are those in which visualization experts engage in the work of the target domain or vice versa. We found immersive data analysis and prototyping activities, as catalogued by the Design by Immersion framework of Hall et al.~\cite{hall2020-immersion}, to be the most fruitful. These correspond to the discover and implement phases of design study methodology. In particular, one visualization expert performed typical analyses ``by hand'' (\autoref{fig:disassembly_artifact}) and both visualization and domain experts engaged in collaborative analyses with the in-process prototype. 

The collaborative analyses provided insight into the data analysis process and feedback on the prototype simultaneously. These analysis sessions occurred during biweekly meetings. The meetings were remote with the lead author driving a pair analytics session, sometimes using the prototype in tandem with ad hoc file browsing when features were not yet implemented or even yet ideated. This process helped us find gaps in our design.

Prototyping the visualization was done in tandem with prototyping the automated analysis. This is also a prototyping activity as noted by Hall et al. and adds a ``moving target'' element as discussed by Williams et al.~\cite{williams2020-moving}. As noted by Williams et al., the copious documentation of tasks and interests over time helped us to prioritize design elements that fulfilled long-standing task-needs over those that had gained attention fleetingly.

As the collaborative, immersive analysis processes required deep attention, we found it especially helpful to have multiple people from the visualization team present when interacting with prototypes. This setup allowed one visualization expert to become fully immersed in the activity and workflow without pause, while reserving another to generate the observation artifacts that were used to refine the task analysis and design over time.

\vspace{2ex}

We note that both of these findings relate to the core stages of the design study framework: discover, design, implement, and deploy~\cite{sedlmair-designstudies-2012}. Design Study Methodology notes that stages may overlap and the process can iterate through any sub-loop of stages. This overlapping and looping describes our workflow with the core stages, to the point where we might even consider them continuous. In particular, the open nature of the {\em discover} phase was important in understanding the differences in needs between CFGExplorer and \projectname as well as adapting to the evolving capabilities of the automated analysis and the refinement naturally arising through iteration.




\section{Conclusion}
\label{sec:conclusion}

We have presented a data abstraction, task analysis, and visual analytics system, \projectname, for analyzing how an application is translated into optimized machine instructions by the compiler. Through evaluation sessions, we showed \projectname assisted in performing tasks common to experts' workflows. We found it was important to support a variety of paths through different representations of the instructions and source code. We also observed that while experts appreciated automated assistance and acknowledged its limitations, integration still led to confusion, which we plan to continue to address in future work.

In conducting this design study, we found that immersive activities such as collaborative analysis sessions, having visualization experts perform analysis workflows, and frequent engagement with unpolished prototypes elicited rich feedback aiding our task analysis and visual design. We also found that despite a high degree of similarity between this design study and a previous one, the transferability between designs was limited. The immersive activities helped us identify this quickly and careful task analysis allowed us to retain the transferable elements.

\section{Acknowledgements}
We thank our study participants for their valuable time and the LLNL LEARN project, LLNS B639881 \& B630670, and NSF III-1656958 \& IIS-1844573 for supporting this research. This work performed under the auspices of the U.S. Department of Energy by Lawrence Livermore National Laboratory under Contract DE-AC52-07NA27344. LLNL-CONF-812737.


\bibliographystyle{abbrv-doi}

\bibliography{main}

\end{document}